\documentclass[preprint,showpacs,showkeywords,preprintnumbers,amsmath,amssymb]{revtex4}

\usepackage{graphicx}% Include figure files
\usepackage{dcolumn}% Align table columns on decimal point
\usepackage{bm}% bold math
\begin{document}

%\title{Entropy of Quantum Fractional Damping System: Another Evidence for the Validity of The Third Law}
\title{Fractional Entropy Decaying and the Third Law of Thermodynamics}

\author{Chun-Yang Wang \footnote{Corresponding author. Electronic mail: wchy@mail.bnu.edu.cn}}
\author{Xue-Mei Zong}
\author{Cui-Feng Sun}
\author{Hong Zhang}
\author{Zhen-Xue Song}

\affiliation{Shandong Provincial Key Laboratory of Laser
Polarization and Information Technology, College of Physics and
Engineering, Qufu Normal University, Qufu 273165, China}

%\date{\today} renormalization group transformation

\begin{abstract}
The quantum thermodynamic property of the fractional damping system is investigated extensively.
A fractional power-law decaying entropy function is revealed which presents another evidence for the validity of the third law of thermodynamics in the quantum dissipative region. Several non-trivial characters are excavated such as that the entropy varies from a non-linear diverging function to a semi-linear decaying function of the fractional exponent as the temperature tends to absolute zero.
\end{abstract}



\maketitle

\section{INTRODUCTION}

Generally in the researching of theoretical physics, nothing is more appreciated than establishing a basic theory or finding a fundamental law.
But after its establishment the most cheering thing is then turns to find more realistic evidence.
During the past few years, as the developing of some attractive fields such as anomalous diffusion \cite{ad01,ad02,ad03}, quantum thermodynamics \cite{ent1} and small systems \cite{sms1}, many basic problems in statistical physics are reconsidered.
For example, experimental study on the thermodynamics of small systems has found that the second law of thermodynamics is violated when the system is driven to a state which is far from equilibrium \cite{2ndl}.
While theoretical studies recently performed on the quantum thermodynamical properties of dissipative systems bring forward several new evidences for the validity of the third law \cite{ent1,PHan,chyw,chy2}.
Rather intriguingly it has been found that strong coupling between the system and reservoir helps to ensure the third law of thermodynamics in the system-plus-reservoir dissipative environment \cite{PHan}.
All these ceaselessly emerging findings keep on encouraging us to make further considerations on the basic theories of statistical physics.

The recent grandly progressing in the subject of fractional Brownian motion (fBm) \cite{fbm1,fbm2,fbm3} dazzled us with brilliancy.
As is indexed by the SCI database, dozens of related papers are published every year in the 1990s;
and in the last five years, it should even be counted by the hundreds.
However abundant may it has been, the low-temperature property of fBms (or systems with certain kind of fractional fluctuating) has not been concerned as far as we have known.
Were it to tell us something different about quantum thermodynamics?
Were there some fundamental laws be obeyed or violated?
All these questions are well worth to be answered.

Therefore in this brief report, we give one of our recent study on the quantum thermodynamical properties of the fractional damping (FD) system which can be understood under the framework of typical fBms.
The free energy and entropy are analytically calculated by using of the remarkable formula \cite{bla,fre1,fre2,fre3}.
The temperature-dependent evolution of them are investigated and compared with the previous studies.
A fast decaying entropy function is witnessed again revealing good conformity with the third law of thermodynamics.

The layout of this study is as follows:
a short introduction of the fBms and fractional damping systems concerning in this report is given in Sec. \ref{sec2};
the detailed information of free energy and entropy calculation is presented in Sec. \ref{sec3};
Section \ref{sec4} serves as a summary of our conclusion where some further discussions are also included.

\section{fractional Brownian motion and fractional damping}\label{sec2}

The extensively concerned fBms in recent studies is the quintessential model for random processes displaying the ``Joseph effect'' (long-range correlations) \cite{jef1}.
Like Brownian motion emerges from the white noise, fBms are generated from the fractional Gaussian noise (fGn) \cite{fbm1,fgn2}
\begin{eqnarray}
\xi(t)=\frac{dx(t)}{dt},
\label{eq:fGn}
\end{eqnarray}
which is zero mean and correlates as $\langle\xi(t_{1})\xi(t_{2})\rangle=2D_{H}H(2H-1)|t_{1}-t_{2}|^{2H-2}$. Where $0<H<1$ is known as the Hurst exponent with $H=1/2$ corresponds to the classical Brownian motion and $H<1/2$ and $H>1/2$ denoting the subdiffusive and superdiffusive cases, respectively.
$D_{H}=[\Gamma(1-2H)\textrm{cos}(H\pi)]/(2H\pi)$ is the diffusion coefficient identified by the Gamma function $\Gamma(z)=\int_{0}^{\infty}t^{z-1}e^{-t}dt$.

Since Eq.(\ref{eq:fGn}) emerges as the force-free overdamped case of the traditional generalized Langevin equation,
the fBms are generally understood as the universal scaling limit of Langevin dynamics whose microscopic-level correlations transcend to the macroscopic level, but yet whose microscopic-level fluctuations do not.
The trajectory sample of fBm is a self-affine stationary Gaussian process and is characterized by a standard normal probability distribution for any $t>0$. i.e.,
\begin{eqnarray}
P(x,t)=\frac{1}{\sqrt{2\pi D_{H}t^{2H}}}\textrm{exp}\left[-\frac{(x-x_{0})^{2}}{4D_{H}t^{2H}}\right]
\label{eq:pdf},
\end{eqnarray}
These equations are very similar to those of classical Brownian motion in the purely dynamical case.

However, in this report, instead of addressing any other point of the fBms,
we make our effort to concern on the low-temperature thermodynamic properties of a typical fractional damping (FD) system which is described by the following operator form of quantum fractional Langevin equation \cite{fle1,fle2,fle3}
\begin{eqnarray}
m\ddot{\hat{x}}+\int^{t}_{0}\eta(t-t')\dot{\hat{x}}(t')dt'=f(\hat{x}(t))+\hat{\xi}(t),
\label{eq:fle}
\end{eqnarray}
which can be derived from a standard Hamiltonian model based on the coupling of Brownian particles to a thermal bath of harmonic oscillators \cite{hotb}. Where $\eta(t)=\eta_{\alpha}t^{-\alpha}/\Gamma(1-\alpha)$ is the particular frictional kernel with exponent $0<\alpha<1$ and constant $\eta_{\alpha}$.
$f(\hat{x}(t))$ is the force function resulting from the external potential $U(\hat{x}(t))$ and $\hat{\xi}(t)$ is a stationary fractional Gaussian noise satisfying the quantum fluctuation-dissipation relation \cite{sdj1,sdj2}
\begin{eqnarray}
\langle \hat{\xi}(t)\hat{\xi}(t')\rangle_{s}=\frac{\beta\hbar}{\pi}\int^{\infty}_{0} d\omega J(\omega)\textrm{coth}(\frac{\beta\hbar\omega}{2})\textrm{cos}(t-t'),
\label{eq:fdr}
\end{eqnarray}
where $\langle \cdots\rangle_{s}$ denotes the quantum symmetric average operation and $\beta=1/k_{B}T$ is the inverse temperature. $J(\omega)\propto\eta_{\alpha}\omega^{\alpha}$ is the spectral bath density corresponding to a sub-Ohmic thermal bath of quasi-infinite
number of oscillators.

\section{entropy and the third law}\label{sec3}

The traditional method for calculating the entropy function of a quantum oscillator is to directly partial differentiate the free energy which can be obtained by using the remarkable formula \cite{bla,fre1,fre2,fre3}
\begin{eqnarray}
F(T)=\frac{1}{\pi}\int^{\infty}_{0}d\omega f(\omega,T)\textrm{Im}\left\{\frac{d\log\Psi(\omega+i0^{+})}{d\omega}\right\}\label{eq:fre},
\end{eqnarray}
where $f(\omega,T)$ is the free energy of a single oscillator of frequency $\omega$, given by $f(\omega,T)=k_{B}T\log[1-\exp(-\hbar\omega/k_{B}T)]$ with the
zero-point contribution $\hbar\omega/2$ being omitted.
While $\Psi(\omega)=\tilde{\hat{x}}(\omega)/\tilde{\hat{\xi}}(\omega)$ denotes the generalized susceptibility which can be got from Eq.(\ref{eq:fle}) by using of the Fourier transformation such as $\tilde{\hat{x}}(\omega)=\int_{-\infty}^{\infty}dt\hat{x}(t)\exp(i\omega t)$.

Since the function $f(\omega,T)$ in Eq. (\ref{eq:fre}) vanishes exponentially for $\omega\gg k_{B}T/\hbar$,
thus as $T\rightarrow0$ the integrand is confined to low frequencies and we can explicitly calculate the free energy and then the entropy by expanding the
factor multiplying $f(\omega,T)$ in the powers of $\omega$.
Supposing the external potential is harmonic: $U(\hat{x})=\frac{1}{2}m\omega^2_0\hat{x}^2$, it can then be inferred from Eq.(\ref{eq:fle}) that
\begin{eqnarray}
\Psi(\omega)=\left[-m\omega^{2}-m\eta_{\alpha}(i\omega)^{\alpha}+m\omega_{0}^{2}\right]^{-1},
\label{eq:alp}
\end{eqnarray}
where a nontrivial part of $(i)^{\alpha}=\textrm{cos}(\alpha\pi/2)+i\textrm{sin}(\alpha\pi/2)$ is contained.
After some algebra, we obtain in the low-frequency limit
\begin{eqnarray}
&&\textrm{Im}\left\{\frac{d\log\Psi(\omega)}{d\omega}\right\}=\frac{\eta_{\alpha}\omega^{\alpha-1}[\alpha(\omega_{0}^{2}-\omega^{2})+2\omega^{2}]
\textrm{sin}\frac{\alpha\pi}{2}}
{\eta_{\alpha}^{2}\omega^{2\alpha}-\eta_{\alpha}\omega^{\alpha}(\omega^{2}-\omega_{0}^{2})\textrm{cos}\frac{\alpha\pi}{2}+(\omega^{2}-\omega_{0}^{2})^{2}}
\nonumber\\
&&\hspace{3.2cm}\cong\frac{\alpha\eta_{\alpha}\omega^{\alpha-1}}{\omega_{0}^{2}}\textrm{sin}(\frac{\alpha\pi}{2})
\label{eq:Im}.
\end{eqnarray}
Thus the free energy is resulted from
\begin{eqnarray}
&&F(T)\cong
\frac{\alpha\eta_{\alpha}k_{B}T}{\pi\omega_{0}^{2}}\textrm{sin}(\frac{\alpha\pi}{2})
\int^{\infty}_{0}d\omega\omega^{\alpha-1}\log[1-e^{-\hbar\omega/k_{B}T}]\nonumber\\
&&\hspace{1.1cm}=-\frac{\alpha\eta_{\alpha}\hbar}{\omega_{0}^{1-\alpha}}\textrm{sin}(\frac{\alpha\pi}{2})\Gamma(\alpha)\zeta(\alpha+1)
\left(\frac{k_{B}T}{\hbar\omega_{0}}\right)^{\alpha+1},
\label{eq:fe}
\end{eqnarray}
where a nontrivial special integral $\int_{0}^{\infty}dyy^{\nu}\textrm{log}(1-e^{-y})=-\Gamma(\nu+1)\zeta(\nu+2)$ is relevant and $\zeta(z)=\Sigma_{n=1}^{\infty}\frac{1}{n^{z}}$ is the Riemann's zeta-function.
The entropy function of the frictional damping system is then can be yielded from
\begin{eqnarray}
&&S(T)=-\frac{\partial F(T)}{\partial T}\nonumber\\
&&\hspace{1.0cm}=\frac{\alpha(\alpha+1)\eta_{\alpha}}{\omega_{0}^{2}(\hbar/k_{B})^{\alpha+1}}\textrm{sin}(\frac{\alpha\pi}{2})\Gamma(\alpha)\zeta(\alpha+1)
T^{\alpha}.
\label{eq:e}
\end{eqnarray}
So we have got as $T\rightarrow0$, $S(T)$ vanishes proportional to $T^{\alpha}$, ensuring again the third law of thermodynamics.

\begin{figure}[ht]
%\centering
\includegraphics[scale=0.7]{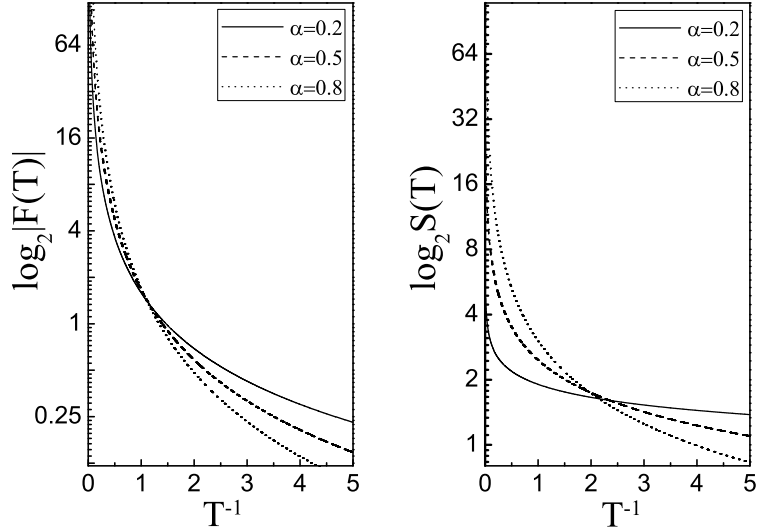}
\caption{Free energy and entropy of the FD system as functions of the inverse temperature where dimensionless parameters such as
$\eta_{\alpha}=\hbar\omega_{_{0}}=k_{B}=1.0$ are used for convenience. \label{fig:f&e}}
\end{figure}

\begin{table}[ht]
\begin{ruledtabular}
\begin{tabular}{lllll}
  Type of QD system                     & $J(\omega)$         & $\textrm{Im}\{{\cal D}(\omega)\}$   &$\quad F(T)$                & $S(T)$        \\
\hline
  harmonic\footnotemark[1]              & \quad $\omega$          &\quad const.                     & \qquad $T^{2}$             & \quad $T$     \\
  Ohmic\footnotemark[2]                 & \quad $\omega$          &\quad const.                     & \qquad $T^{2}$             & \quad $T$     \\
  Drude\footnotemark[3]                 & \quad $\omega$          &\quad const.                     & \qquad $T^{2}$             & \quad $T$     \\
  harmonic velocity\footnotemark[1]     & \quad $\omega^{3}$      &\quad$\omega$                    & \qquad $T^{3}$             & \quad $T^{2}$ \\
  blackbody radiation\footnotemark[2]   & \quad $\omega^{3}$      &\quad$\omega^{2}$                & \qquad $T^{4}$             & \quad $T^{3}$ \\
  harmonic acceleration\footnotemark[1] & \quad $\omega^{5}$      &\quad$\omega^{3}$                & \qquad $T^{5}$             & \quad $T^{4}$ \\
  non-Ohmic\footnotemark[3]             & \quad $\omega^{\delta}$ &\quad$\omega^{\delta-1}$         & \qquad $T^{\delta+1}$      & \quad $T^{\delta}$ \\
  fractional damping                    & \quad $\omega^{\alpha}$ &\quad$\omega^{\alpha-1}$         & \qquad $T^{\alpha+1}$      & \quad $T^{\alpha}$ \\
  arbitrary chosen bath\footnotemark[2]             & \quad $\omega^{\nu+1}$  &\quad$\omega^{\nu}$              & \qquad $T^{\nu+2}$         & \quad $T^{\nu+1}$  \\
\end{tabular}
\end{ruledtabular}
\footnotetext[1]{Ref.\cite{chyw}.}
\footnotetext[2]{Ref.\cite{ent1}, where $\nu\in(-1,1)$.}
\footnotetext[3]{Ref.\cite{chy2}, where $\delta\in(0,2)$.}
\caption{Comparison of the thermodynamic properties of variant conventional QD systems with different spectra density $J(\omega)$. Replacement ${\cal D}(\omega)=d\log\Psi(\omega)/d\omega$ is used for simplicity.}
\label{tab:t1}
\end{table}

\begin{figure}[ht]
%\centering
\includegraphics[scale=0.9]{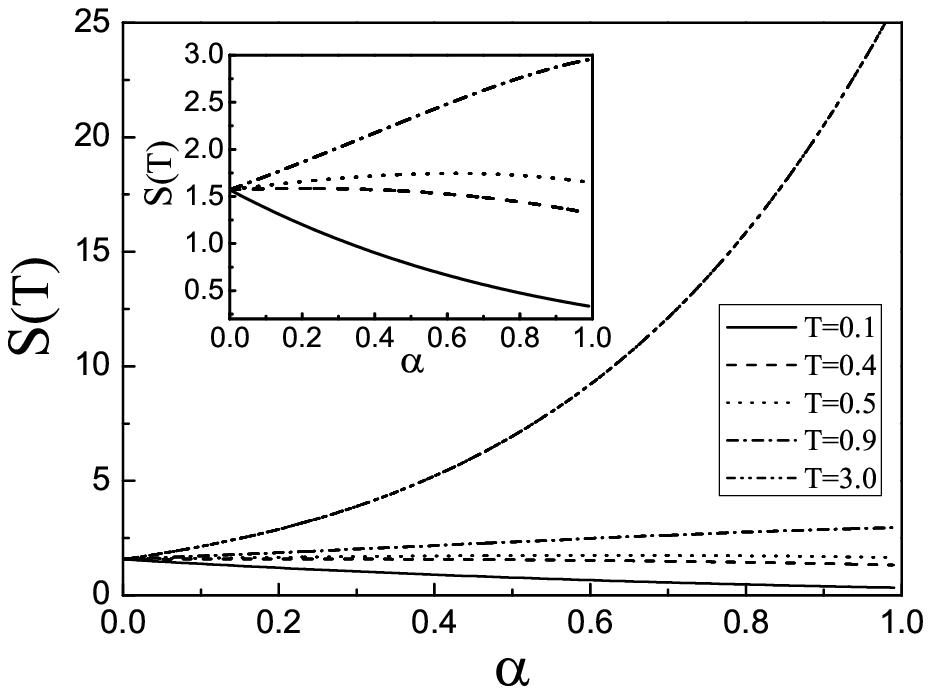}
\caption{Entropy of the FD system plotted as a function of the fractional exponent at variant system temperature. Inserted are the local amplification of the linear curves, Dimensionless units are used as those of Fig.\ref{fig:f&e}. \label{fig:e&a}}
\end{figure}

In order to explicitly illustrate the temperature-dependent decaying behaviour of the functions, we plot in Fig.\ref{fig:f&e} the free energy and entropy of the FD system as functions of the inverse temperature at variant cases of $\alpha$.
From which we can see that in all cases the functions decay rapidly and the decaying rate increases as the increasing of $\alpha$.
This is reminiscent of the previous studies.
But noticing that $\alpha$ is the fractional exponent of the friction kennel function.
So this emerges a bit non-trivial. One may argue that is it a mere coincidence or an inevitable consequence for the FD system?

For a fact-clarification, we give in Table.\ref{tab:t1} a comparison about the thermodynamic property of variant conventional quantum dissipative (QD) systems.
The temperature-dependent relation of $F(T)$ and $S(T)$ are listed as well as the asymptotically frequency-depending of the power spectra density $J(\omega)$.
Where we can find that the way of entropy-decaying for the FD system is identical to that of the non-Ohmic system except for the definition domain of the power exponent.
The reason that lead to these similarity may lives in the identical frequency-dependent form of their spectra density.
In fact, supposing the asymptotic spectra density of a system can be identified by a universal power exponent, then we can find in Table.\ref{tab:t1} that there lives a very similar trend of power law decaying in all the conventional QD systems whose power exponent is no large than 2 (no matter it is $\alpha$, $\delta$ or $\nu$ and even others).
Therefore, in our opinion, the fractional entropy decaying of the FD system is revealed to be an inevitable consequence under the quantum thermodynamic circumstance.

For more in-depth understanding, we give in Fig.\ref{fig:e&a} a further investigation on the $\alpha$-dependent varying of the fractional decaying entropy $S(T)$ at different values of system temperatures.
From which we can see that, as the temperature goes down, $S(T)$ varies from a non-linear diverging function to a semi-linear decaying function of $\alpha$.
In the low-temperature limit it decays almost standardly in a linear form.
This is a non-trivial phenomena which has never been witnessed before.
Although the intrinsic mechanism leading to this results is not clear in the current background of knowledge, one can still believe that there lives numerous novelties to be revealed in the FD systems.

\section{summary and discussion}\label{sec4}

In conclusion, we have presented in this report an extensive analysis on the quantum thermodynamic properties of the fractional damping systems.
The free energy and entropy functions are calculated which is revealed to have a fractional power-law decaying character as the temperature tend to absolute zero. This presents another evidence for the validity of the third law of thermodynamics in the quantum dissipative region.
Moreover, some non-trivial phenomena is found such as that $S(T)$ varies from a non-linear diverging function to a semi-linear decaying function of $\alpha$ as the temperature goes down. All these findings have revealed the innumerable not excavated novelties relevant to the fBm and FD systems.

Experimentally a growing body of single-trajectory studies have suggested that, among the variety of stochastic processes that produce sub-diffusion, fBm may be a model particularly relevant to subcellular transport.
For example, the negatively and long-range correlation emerging when $H<1/2$ were observed in sub-diffusing mRNA molecules \cite{sd10}, RNA proteins, and chromosomal loci within E. coli cells\cite{sd04}.
Similarly, fBm can be used to describe unbiased translocations \cite{sd26,sd27}, the dispersion of apoferritin proteins in crowded dextran solutions \cite{sd11} and lipid molecules in lipid bilayers \cite{sd12}.
Therefore, although several aspects for the understanding of fBm and FD systems remain formidable, we expect that this work would contribute toward the continuous demystification of this seemingly simple subject.

\section * {ACKNOWLEDGEMENTS}

This work was supported by by the Shandong Province Science Foundation for
Youths under the Grant No.ZR2011AQ016.

\end{document}